\documentclass[manuscript]{aastex}
\usepackage{spr-astr-addons}

\begin{document}
\title{Compact star in pseudo-spheroidal spacetime}

\author{D. Shee\altaffilmark{1}, S. Ghosh\altaffilmark{2}, F. Rahaman\altaffilmark{3}, B.K. Guha\altaffilmark{4}, Saibal Ray\altaffilmark{5}}
\altaffiltext{1}{Department of Physics, Indian Institute of
Engineering Science and Technology, Shibpur, Howrah 711103, West
Bengal, India,\\dibyendu\_shee@yahoo.com}

\altaffiltext{2}{Department of Physics, Indian Institute of
Engineering Science and Technology, Shibpur, Howrah 711103, West
Bengal, India,\\shnkghosh122@gmail.com}

\altaffiltext{3}{Department of Mathematics, Jadavpur University,
Kolkata 700032, West Bengal, India,\\rahaman@associates.iucaa.in}

\altaffiltext{4}{Department of Physics, Indian Institute of
Engineering Science and Technology, Shibpur, Howrah 711103, West
Bengal, India,\\bkguhaphys@gmail.com}

\altaffiltext{5}{Department of Physics, Government College of
Engineering \& Ceramic Technology, Kolkata 700010, West Bengal,
India,\\saibal@associates.iucaa.in}

\begin{abstract}
We investigate perfect fluid stars in $(2+1)$ dimension
 in pseudo-spheroidal spacetime with the help of Vaidya-Tikekar
 metric where the physical $3$-space ($t=$ constant) is described by pseudo-spheroidal
 geometry. Here the spheroidicity parameter $a$, plays an important role
 for determining the properties of a compact star. In the present work
 a class of interior solutions corresponding to the Ba$\tilde{n}$ados-Teitelboim-Zanelli $(BTZ)$ (Bañados et al., Phys.
Rev. Lett. 69:1849, 1992) exterior metric has been provided which describes a static circularly symmetric star with
 negative cosmological constant in equilibrium. It is shown that asymptotically
 anti-de Sitter $(2+1)$ dimensional spacetime described by BTZ admits a compact star
 solution with reasonable physical features.
\end{abstract}

\keywords{General relativity; Pseudo-spheroidal spacetime; Compact star}

\section{Introduction}
After the discovery of the particle named as Neutron by Chadwick,
later on existence of the neutron star was predicted. A neutron
star is the later stage of a gravitationally collapsed star. It
becomes stabilized by the degenerate neutron pressure, after
exhausting all its thermonuclear fuel. With the discovery of
pulsars~\citep{Hewish1968}
 this concept got concrete experimental support. However the estimated mass and radius of
 different compact object such as X-ray pulsar $Her~X-1$, X-ray burster $4U~1820-30$,
  millisecond pulsar $SAX~J~1808.4-3658$ and X-ray sources $RX~J~185635-3754$
could not be described by the standard neutron star model. It is
found by~\citet{Ruderman1972} that the matter densities
of compact stars are to be of the order of $10^{15}$ gm/cc or
higher, exceeding the nuclear matter density and at this high
density range nuclear interactions must be treated
relativistically. As a result of the anisotropy, pressure inside
the fluid sphere can be decomposed into two parts namely radial
pressure $p_r$ and the transverse pressure $p_t$, where $p_t$ is
in the perpendicular direction to $p_r$. Anisotropy may occurs in
various reasons e.g. the existence of solid core, in presence of
type P superfluid, phase transition, rotation, magnetic field,
mixture of two fluid, existence of external field
etc~\citep{Thirukkanesh2014}. On the basis of compactification
factor (ratio of mass and radius), the compact objects are
classified into a normal star ($10^{-5}$), white dwarf
($10^{-3}$), neutron star ($0.1$ to $0.2$), strange star ($0.2$
to $<0.5$), black hole ($0.5$) etc. The physics and equation of
state (EOS) of compact objects near the core region are still not
known clearly.

To analyze a compact object~\citet{Vaidya1982} and~\citet{Tikekar1990} prescribed a simple form
for the space like hyper surface ($t$= constant) containing two
parameters, namely spheroidicity parameter $(a)$ and curvature
parameter ($R$). The Vaidya-Tikekar approach reduces the
complexity of field equations which produces solution of
relativistic stars with ultra high densities and is also useful to
obtain stellar solution for a compact star with Einstein-Maxwell
field equations~\citep{Rhodes1974}. The physics of the
Vaidya-Tikekar metric is discussed in the
Refs.~\citep{Vaidya1982,Knutsen1988,Tikekar1990}. With different
choice of spheroidal parameter~\citet{Maharaja1996} obtained a new classes of solutions for
superdense stars.~\citet{Thomas1998} and~\citet{Tikekar1999} obtained a class of
relativistic solutions analyzing compact stars with
$3$-pseudo-spheroidal geometry for the $3$-space of the interior
spacetime. Considering the anisotropic distribution of fluid
including pseudo-spheroidal geometry to construct compact star
models one may look at the Refs.~\citep{Patel1995,Jotania2005}. In
connection to compact star in one of our earlier
works~\citep{Shee2016} we proposed a model for relativistic dense
star with anisotropy that admits non-static conformal symmetry.

Actively researching in the field of lower dimensional gravity one
can understand substantially some of the crucial points of
astrophysics. Lower dimensional analysis of black holes has often
been preferred to understand the various issues which are to be
difficult to resolve in conventional dimensions.
~\citet{Zanelli1992} (henceforth
BTZ) obtained a beautiful solution which
opened up the possibility of investigating many interesting
features of black holes. They obtained a analytical solution
representing the exterior gravitational field of a black hole in
$(2+1)$ dimensions in the presence of negative cosmological
constant.~\citet{Mann1993} analyzed collapsing dust
cloud ($p=0$) in $(2+1)$ dimensions leading to a black hole. Later
on~\citet{Martins2010} obtained a self similar
solution in $(2+1)$ dimensions by considering the collapse of a
circularly symmetric anisotropic fluid. The interior solution for
an incompressible fluid in $(2+1)$ dimensions and the bound on the
maximum allowed mass of the resulting configuration was obtained
by~\citet{Cruz1995}. The study also claims that
the collapsed stage would always be covered under its event
horizon. A class of interior solutions corresponding to BTZ
exterior was provided by~\citet{Cruz2005} by assuming
a particular density profile $\rho=\rho_c(1-\frac{r^2}{R^2})$,
where $\rho_c$ is the central density, $\rho$ is the density which
is a function of the radial parameter $r$ and $R$ is the boundary
of the star. Assuming polytropic EOS, such as
$p=K\rho^{1+\frac{1}{n}}$,~\citet{Paulo1999} proposed a
interior solution corresponding to the BTZ exterior, where $n$ is
the polytropic index and $K$ is the polytropic constant. A new
class of interior solution corresponding to the BTZ exterior was
provided by~\citet{Sharma2011} by assuming a particular
form of mass function $2m(r)=C-e^{-2\mu(r)}-\Lambda r^2$, where
$\mu(r)$ is the metric function and $m(r)$ is the mass within the
radial distance. The
general BTZ metric is characterized by its mass, angular momentum
and electric charge but is asymptotically anti-de Sitter rather
than flat~\citep{Husain1995}.

Being motivated by the above background works we have presented here a
compact star model under ($2+1$) dimensional metric with several
interesting physical properties. The plan of the investigation is as follows:
In the Sec. 2 we have provided necessary spacetime background and hence the
Einstein field equations in the presence of cosmological constant.
We have found solutions for different physical parameters and matching condition
in the Secs. 3 and 4 respectively. In Sec. 5 we have
explored different physical features, viz. the density and mass,
pressure and anisotropy, stability, energy conditions, compactness and redshift etc. with elaborate discussion.
In the last Sec. 6 we have made some concluding remarks regarding
different aspects of the present model.

\section{The spacetime metric}

\subsection{Interior spacetime}
We take the following ($2+1$) dimensional metric describing the
interior of a static spherically symmetric distribution of matter as
\begin{equation}
ds^2 = -e^{2\gamma(r)} dt^2 + e^{2\mu(r)} dr^2 + r^2d\theta^2.
\label{eq1}
\end{equation}
The energy-momentum tensor of the matter distribution in the
interior of the star is given by
\begin{equation}
T_{ij} = (\rho + p_r) u_i u_j + p_t g_{ij} + (p_r -
p_t)\chi_i \chi_j, \label{eq2}
\end{equation}
where $\rho$ represents the energy density, $p_r$ is the radial pressure,
$p_t$ is the tangential pressure, $g_{ij}$ are the metric tensors, $\chi^{i} = e^{-\mu(r)}\delta^i_r$
is a unit three vector along the radial direction, and $u^{i}$ are the
$3$-velocity of the fluid.

The Einstein field equations with a cosmological
constant under the specification $G = c = 1$ are given by
\begin{eqnarray}
2\pi \rho +\Lambda &=& \frac{\mu' e^{-2\mu}}{r}, \label{eq3} \\
2\pi p_r  -\Lambda &=& \frac{\gamma' e^{-2\mu}}{r}, \label{eq4}
\\ 2\pi p_t  -\Lambda &=&
e^{-2\mu}\left(\gamma'^2+\gamma''-\gamma'\mu'\right),\label{eq5}
\end{eqnarray}
where a `$\prime$' denotes differentiation with respect to the radial parameter $r$.
 Combining Eqs.~(\ref{eq3})-(\ref{eq5}), we have
\begin{equation}
\left(\rho + p_r\right)\gamma' + p_r' + \frac{1}{r}\left(p_r -
p_t\right) =0, \label{eq6}
\end{equation}
which is the conservation equation in $(2+1)$ dimensions.

\subsection {Metric potential}
Since we take $(2+1)$ dimensions in pseudo spheroidal spacetime,
therefore, we use the ansatz~\citep{Mann1993}
\begin{equation}
e^{2\mu} = \frac{1 +\frac{ar^2}{R^2}}{1 +\frac{r^2}{R^2}},\label{eq7}
\end{equation}
where $a$ is the spheroidicity parameter and $R$ is a geometrical parameter
related with the configuration of the star model.

Using Eq.~(\ref{eq7}) we get
\begin{equation}
2\mu=ln\frac{R^2+ar^2}{R^2+r^2}. \label{eq8}
\end{equation}

Therefore differentiating $\mu$ with respect to $r$ we get
\begin{equation}
\mu'=\frac{R^2r(a-1)}{(R^2+ar^2)(R^2+r^2)}. \label{eq9}
\end{equation}

The above equations are used to calculate $\rho$ and $p_t$.
However, here it is to note that the spheroidicity parameter $a$ and curvature parameter $R$
plays a very important role in our work. We shall take $a=6$ and
$R=22.882$ km through out the work as taken by~\citet{Chattopadhyay2012} for X ray pulsar $Her-X-1$. It can be observed that the variation
of these parametric values within the range $5<a<10$ and $20<R<25$
shows very negligible effect. However, for the present prescription on the numerical values of
$a$ and $R$, we would add that though BTZ black hole is not a three-dimensional section of a
$4D$ black hole however it facilitates to understand several key features of the model presented here.

\section {The anisotropic stellar model}
It is known that $\Lambda>0$ implies the space is open. To explain
the present accelerating state of the universe, it is believed
that vacuum energy is responsible for this expansion. As a
consequence, it provides the gravitational effect on
the stellar structures and this cosmological constant $(\Lambda)$
plays the role of vacuum energy or dark energy. In this section we will
study the following features of our model assuming the value of
$\Lambda=0.00018 \ km^{-2}$~\citep{Kalam2012}. We have assumed this value as required
for the stability of the compact star and mathematical
consistency.

The matter density $(\rho)$ can be found from Eq.~(\ref{eq3}) as
\begin{equation}
\rho=\frac{1}{2\pi}\frac{R^2(a-1)}{(R^2+ar^2)^2}
-\frac{\Lambda}{2\pi}. \label{eq10}
\end{equation}

The variation of $\rho$ i.e. matter density with distance from the
center of the star is given by
\begin{equation}
\rho'=\frac{d\rho}{dr}=-\frac{4arR^2(a-1)}{2\pi(R^2+ar^2)^3}<0. \label{eq11}
\end{equation}

The above expression implies that at $r=0$ the matter density remains constant.
The second order derivative of $\rho$ with respect to the distance from
the center of the star is given by
\begin{equation}
\rho''=\frac{d^2\rho}{dr^2}=-\frac{4aR^2(a-1)(R^2-5ar^2)}{2\pi(R^2+ar^2)^4}<0. \label{eq12}
\end{equation}

The variation of $\rho$, $\rho{'}$ and $\rho{''}$ with the radial
distance $r$ are shown in the Fig. 1.

\begin{figure*}[thbp]
\centering
\includegraphics[width=0.35\textwidth]{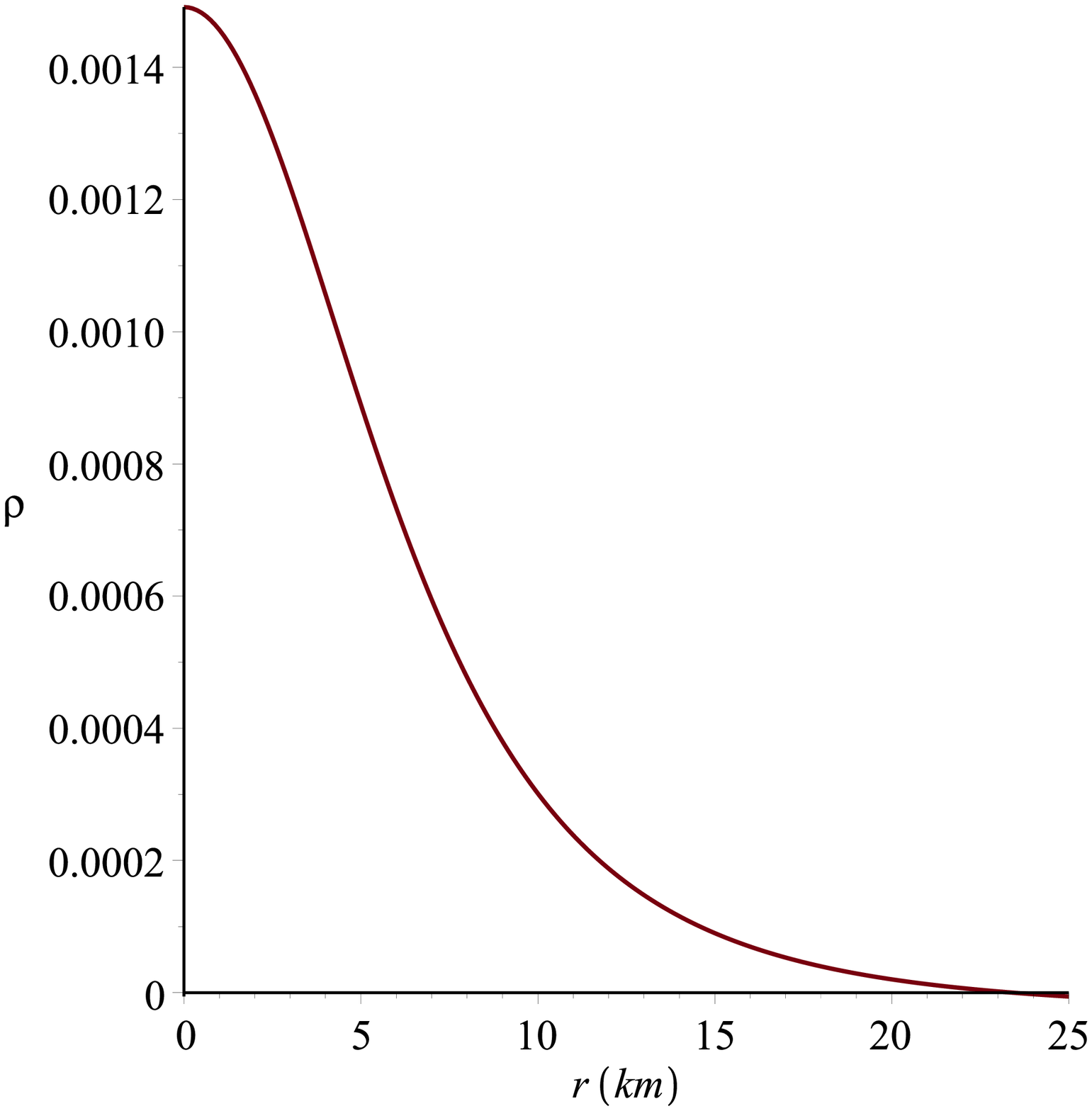}
\includegraphics[width=0.35\textwidth]{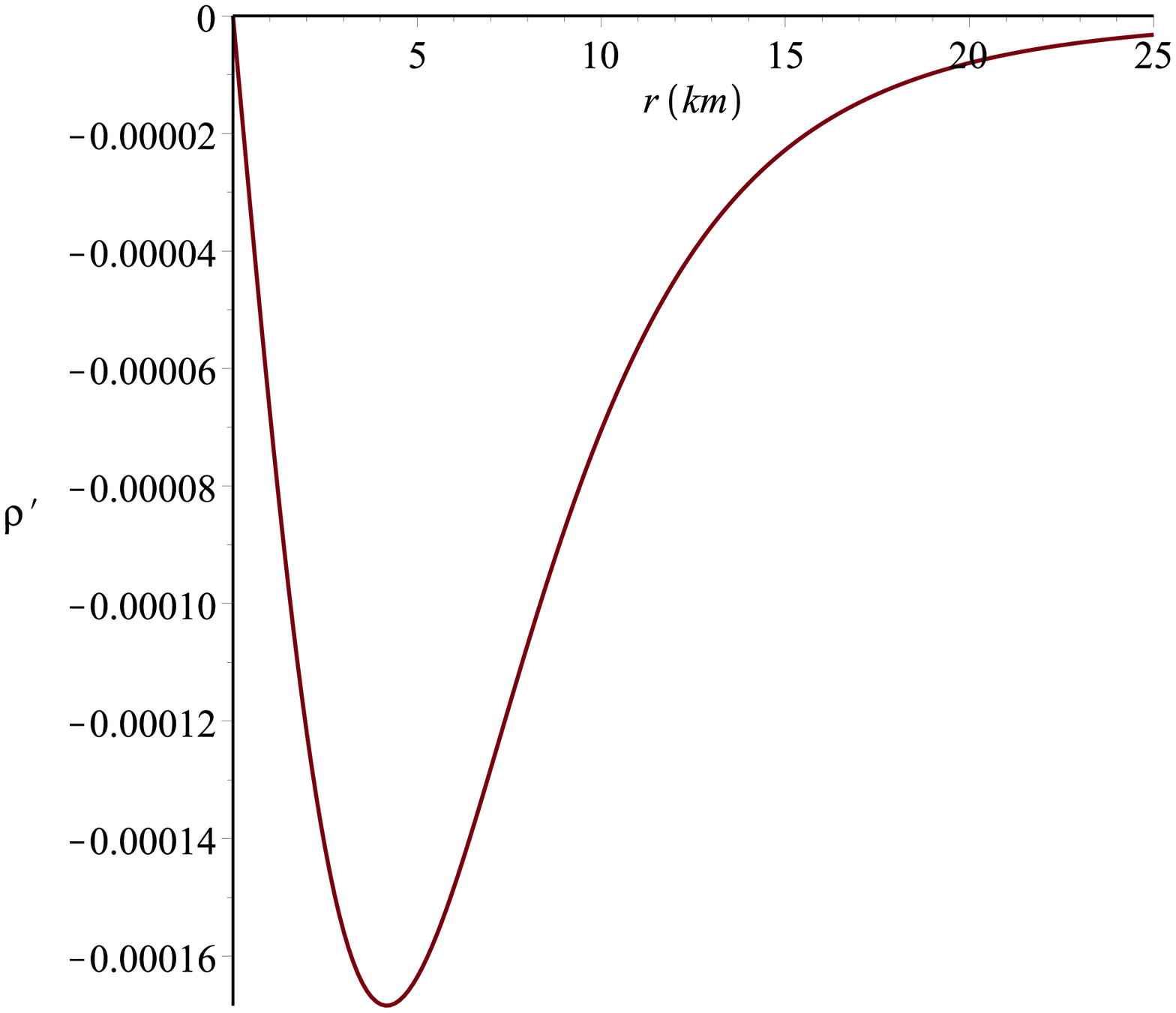}
\includegraphics[width=0.35\textwidth]{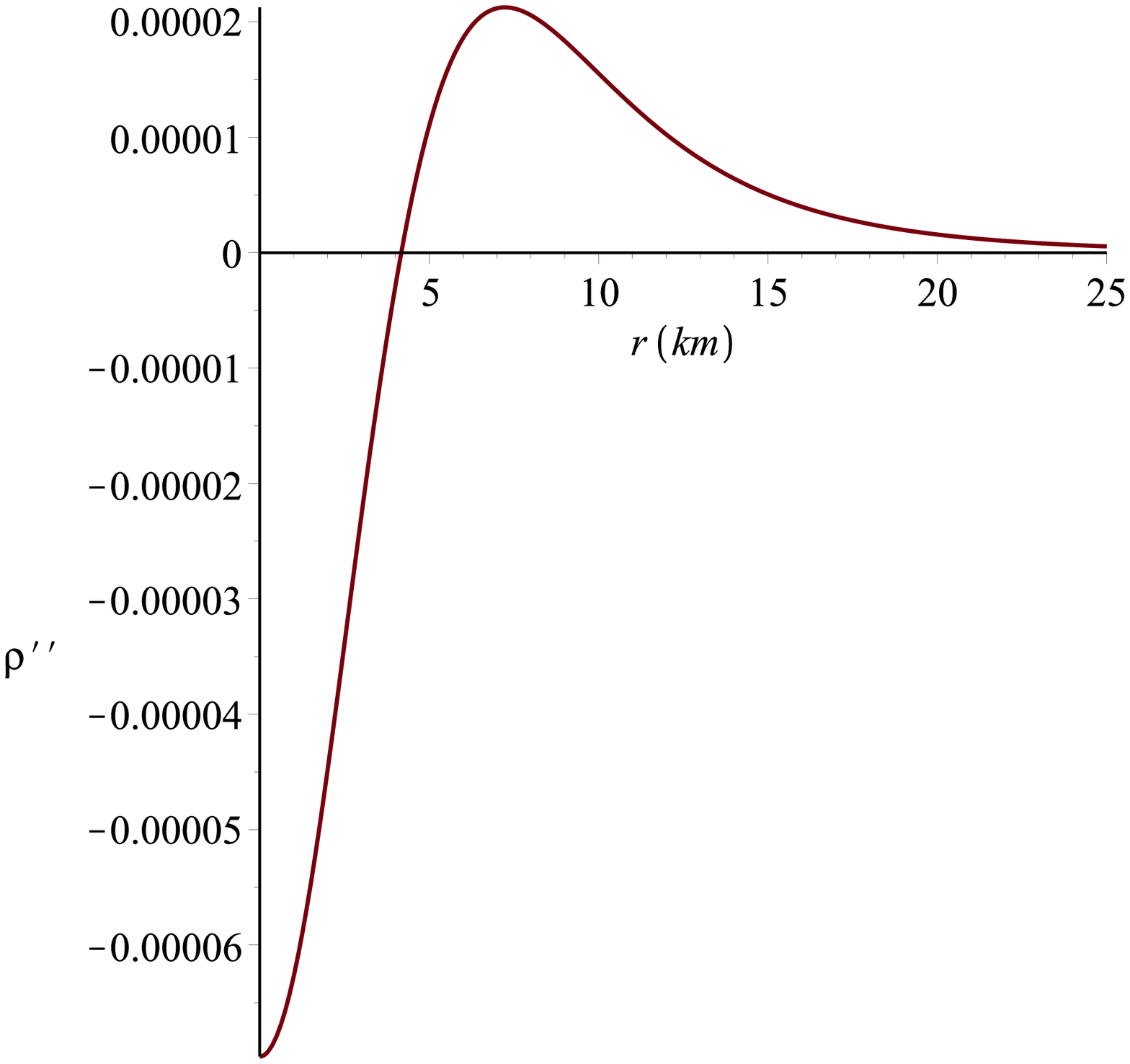}
\caption{Variation of the $\rho$ (upper left), $\rho^{'}$ (upper right) and $\rho^{''}$ (lower) with $r$}
\end{figure*}

If we take $ p_r= \omega \rho $ then we get the differential
equation
\begin{eqnarray}
\gamma'=\frac{r}{(R^2+ar^2)(R^2+r^2)}\left[\omega (a-1)R^2- \right. \nonumber \\ \left. \Lambda (\omega +1)(R^2+ar^2)^2\right]. \label{eq13}
\end{eqnarray}

Integrating the above equation we can get $\gamma$ as
\begin{equation}
\gamma=\frac{\omega}{2}\ln \left(\frac{R^2+ar^2}{R^2+r^2}\right)-\frac{\Lambda}{2}(\omega+1)[ar^2-R^2 \ln
a]+D, \label{eq14}
\end{equation}
where $D$ is the integration constant which can be determined from
boundary conditions.

Using Eqs.~(\ref{eq7}) and ~(\ref{eq13}) we can get the radial pressure $(p_r)$ from
Eq.~(\ref{eq4}) as
\begin{equation}
2\pi p_r=\frac{\omega (a-1)R^2}{(R^2+ar^2)^2}-\Lambda \omega. \label{eq15}
\end{equation}

\begin{figure*}[thbp]
\centering
\includegraphics[width=0.35\textwidth]{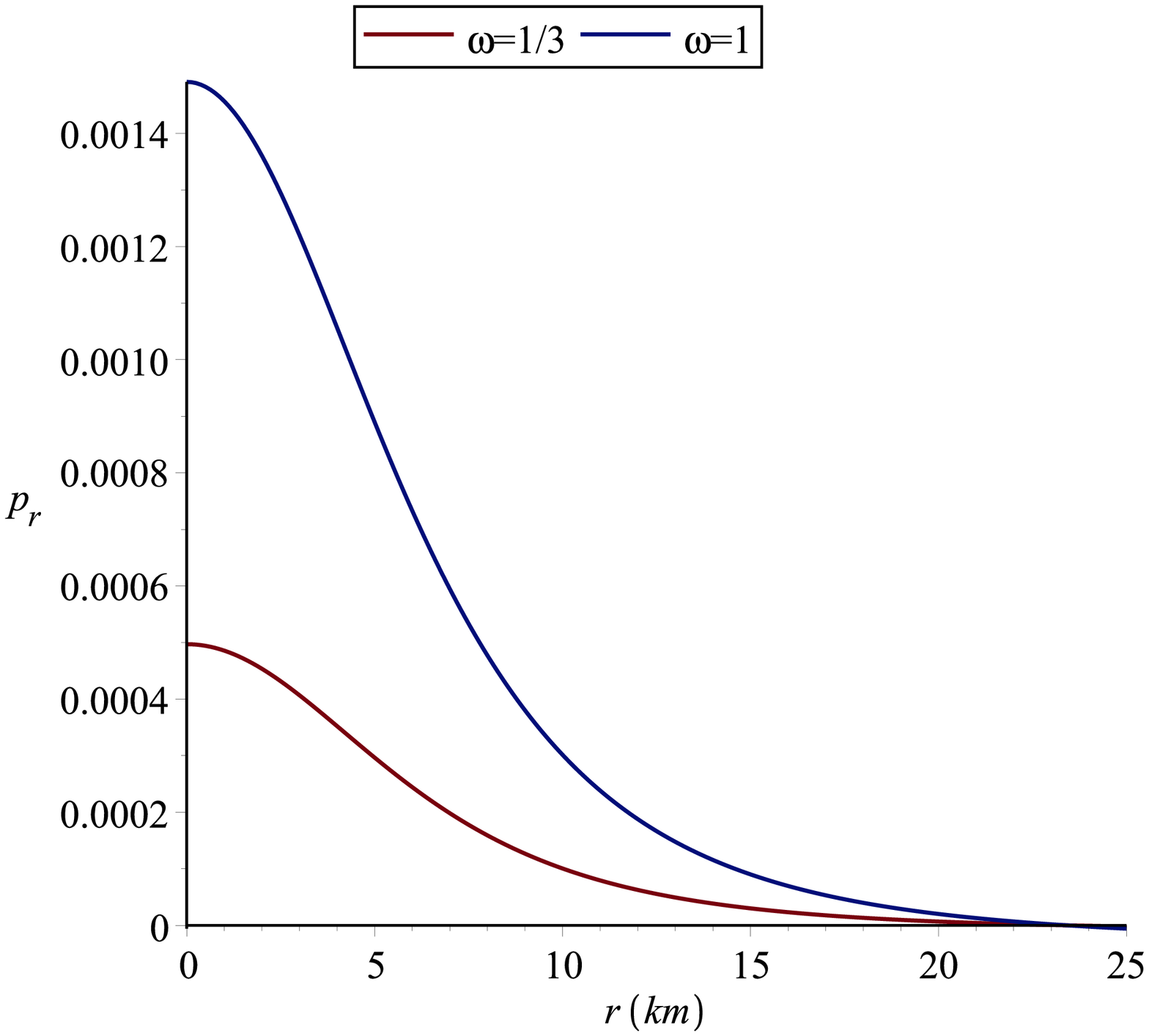}
\includegraphics[width=0.35\textwidth]{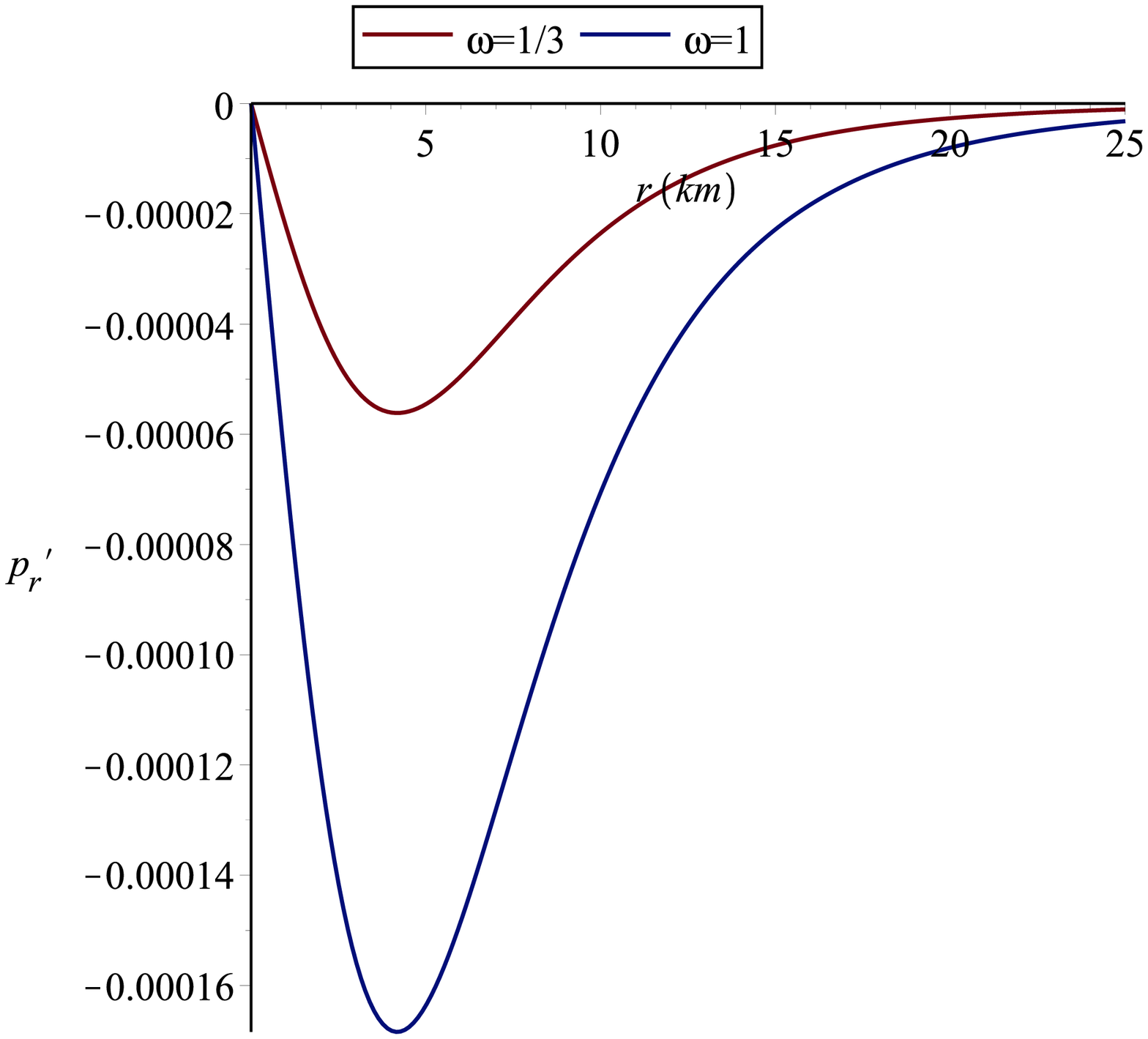}
\includegraphics[width=0.35\textwidth]{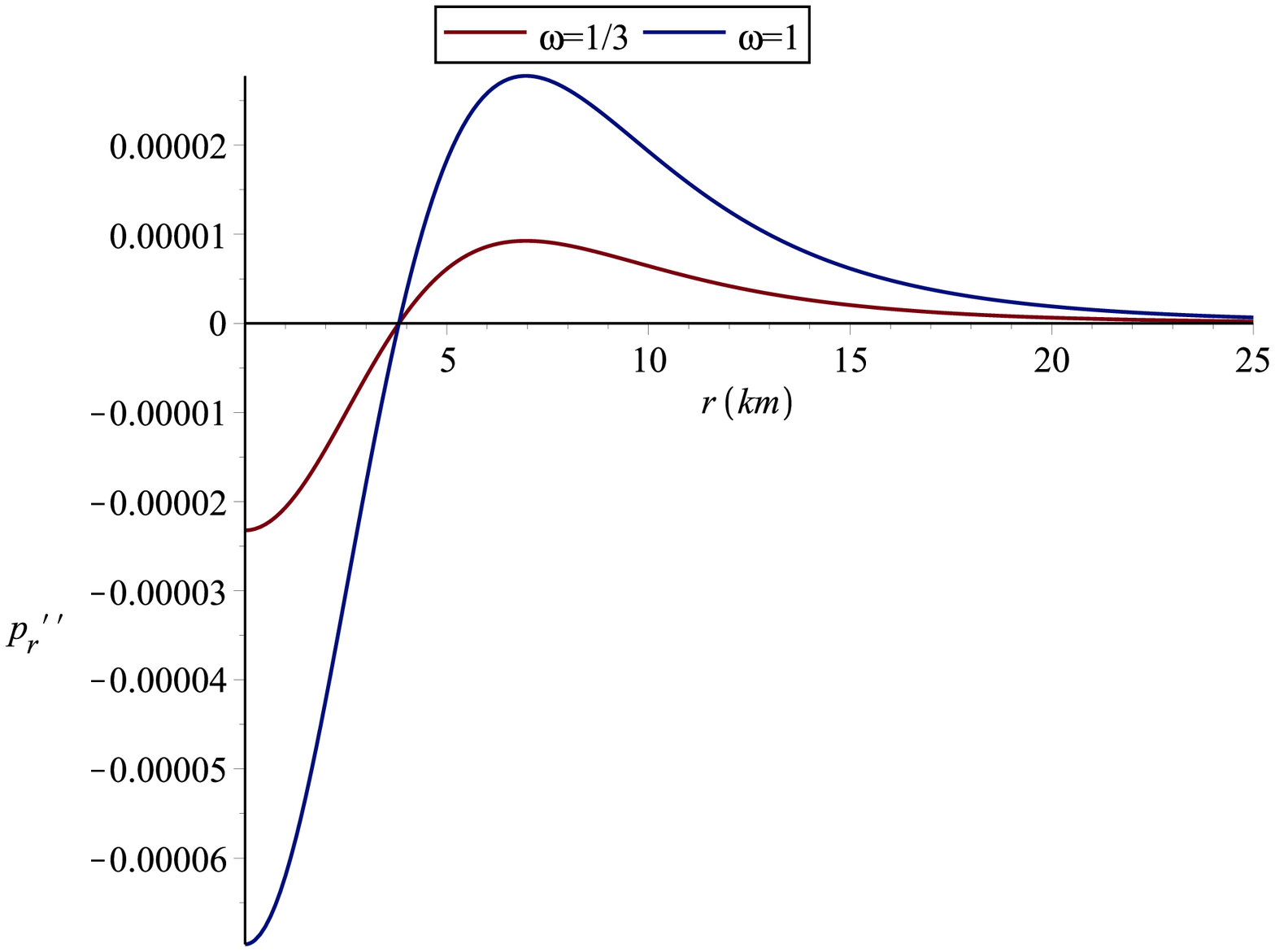}
\caption{Variation of the $p_r$ (upper left), $p_r^{'}$ (upper right) and $p_r^{''}$ (lower) with $r$}
\end{figure*}

Now we shall calculate variation of the radial pressure $(p_r)$ with the
radius $(r)$ of the star as
\begin{equation}
2\pi \frac{dp_r}{dr}=-\frac{4\omega arR^2(a-1)}{(R^2+ar^2)^3}<0. \label{eq16}
\end{equation}

From the above Eq.~(\ref{eq15}) we have the radial pressure $p_r$ is
constant at $r=0$. The second order derivative of radial pressure with
respect to radius of the star is given by
\begin{equation}
2\pi \frac{d^2p_r}{dr^2}=-\frac{4\omega
aR^2(R^2-6ar^2)(a-1)}{(R^2+ar^2)^4}<0. \label{eq17}
\end{equation}

The variation of $p_r$, $p_r{'}$ and $p_r{''}$ with the radial
distance $r$ are shown in the Fig. 2. Since $\frac{d\rho}{dr},\frac{dp_r}{dr}<0$ we can
conclude that both the matter density $(\rho)$ and radial pressure $(p_r)$ are
monotonic decreasing function of radius $r$. From Figs. 1 and 2 it can be observed
that these parameters have maximum value at the center $(r=0)$ of the star and it decreases
 radially outward.

The variation of the radial pressure with respect to the matter density is
given by
\begin{equation}
\frac{dp_r}{d\rho}=\omega=v_{rs}^2, \label{eq18}
\end{equation}
and has been plotted in Fig. 6 which gives a constant value. This result physically implies that
the radial pressure dose not changes with the matter density.

The EOS parameter corresponding to radial direction may be written as
\begin{equation}
\omega_r=\frac{p_r}{\rho}=\omega, \label{eq19}
\end{equation}
which is a constant quantity.

As $p_r=0$ and $\rho=0$ at $r=b$, the radius of compact star, we
get the same result from Eqs.~(\ref{eq10}) and ~(\ref{eq15}) as
\begin{equation}
b=\sqrt{\frac{R}{a}\left(\sqrt \frac{a-1}{\Lambda}-R\right)}. \label{eq20}
\end{equation}

Putting the numerical values of the constants we get the radius of
the star to be $23.417$ km which is exactly same as obtained from Figs. 1 and 2.

The tangential pressure is given by the following equation
\begin{eqnarray}
2\pi p_t-\Lambda =\frac{1}{C^3(R^2+r^2)} \left\{r^2\left[ (A R^2-B
C^2)\right] \right. \nonumber\\ \left. \times \left[R^2(A-a+1)-B C^2\right] \right. \nonumber\\ \left. + A
R^2[R^4-R^2r^2(a+1)-3ar^4] - \right. \nonumber\\ \left. BC[R^4+R^2r^2(3a-1)+ar^4]\right\},\label{eq21}
\end{eqnarray}
where $A=\omega(a-1)$, $B=\Lambda(\omega+1)$ and $C=R^2+ar^2$. The above equation
shows that at $r=0$ the tangential pressure $p_t$ has a
finite positive value. Due to complexity in the expression of
$p_t{'}$ and $p_t{''}$ are not given but their graphical variations
are shown here. The variation of $p_t$, $p_t{'}$ and $p_t{''}$
with the radial distance r are shown in the Fig. 3.

\begin{figure*}[thbp]
\centering
\includegraphics[width=0.35\textwidth]{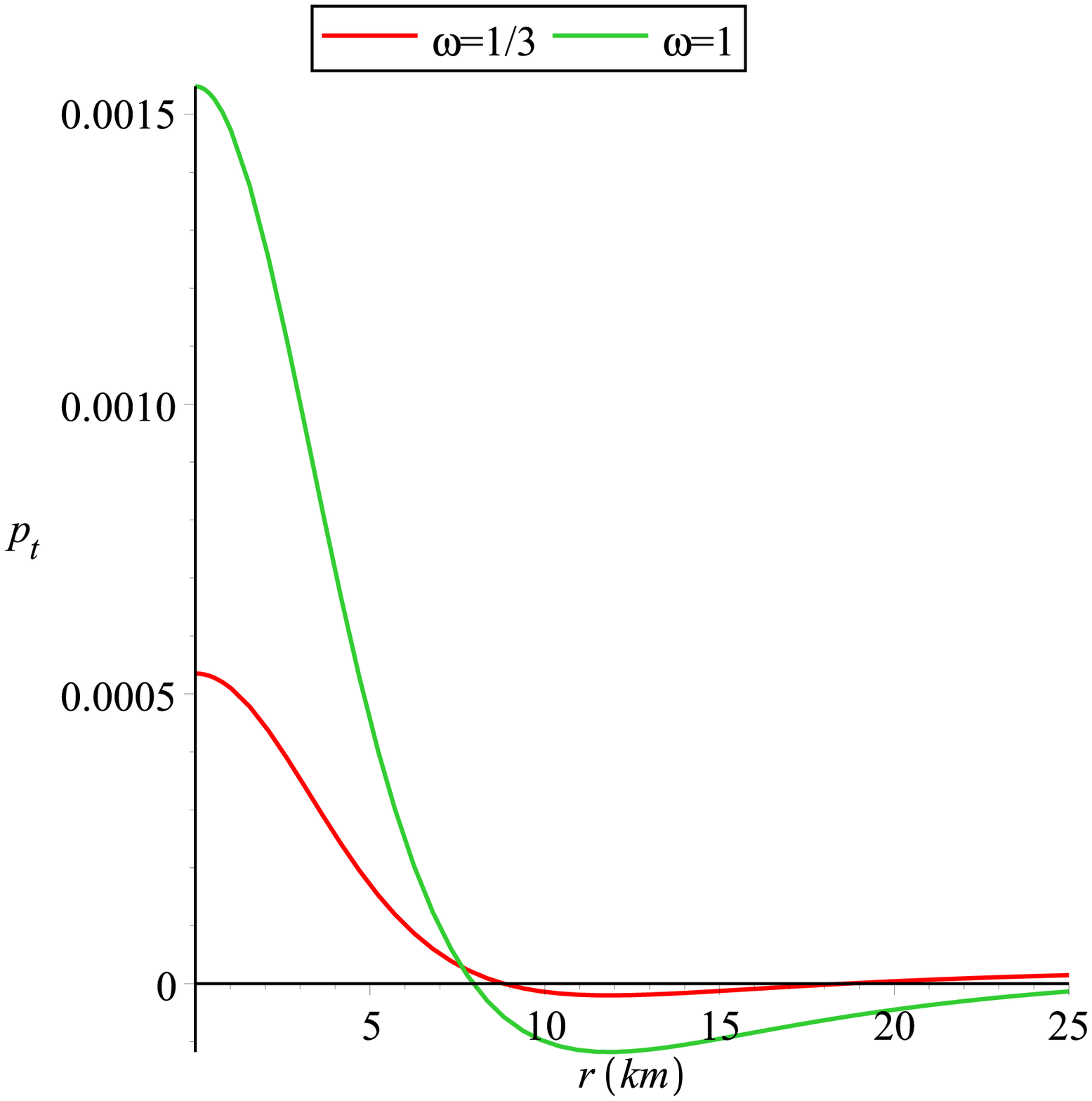}
\includegraphics[width=0.35\textwidth]{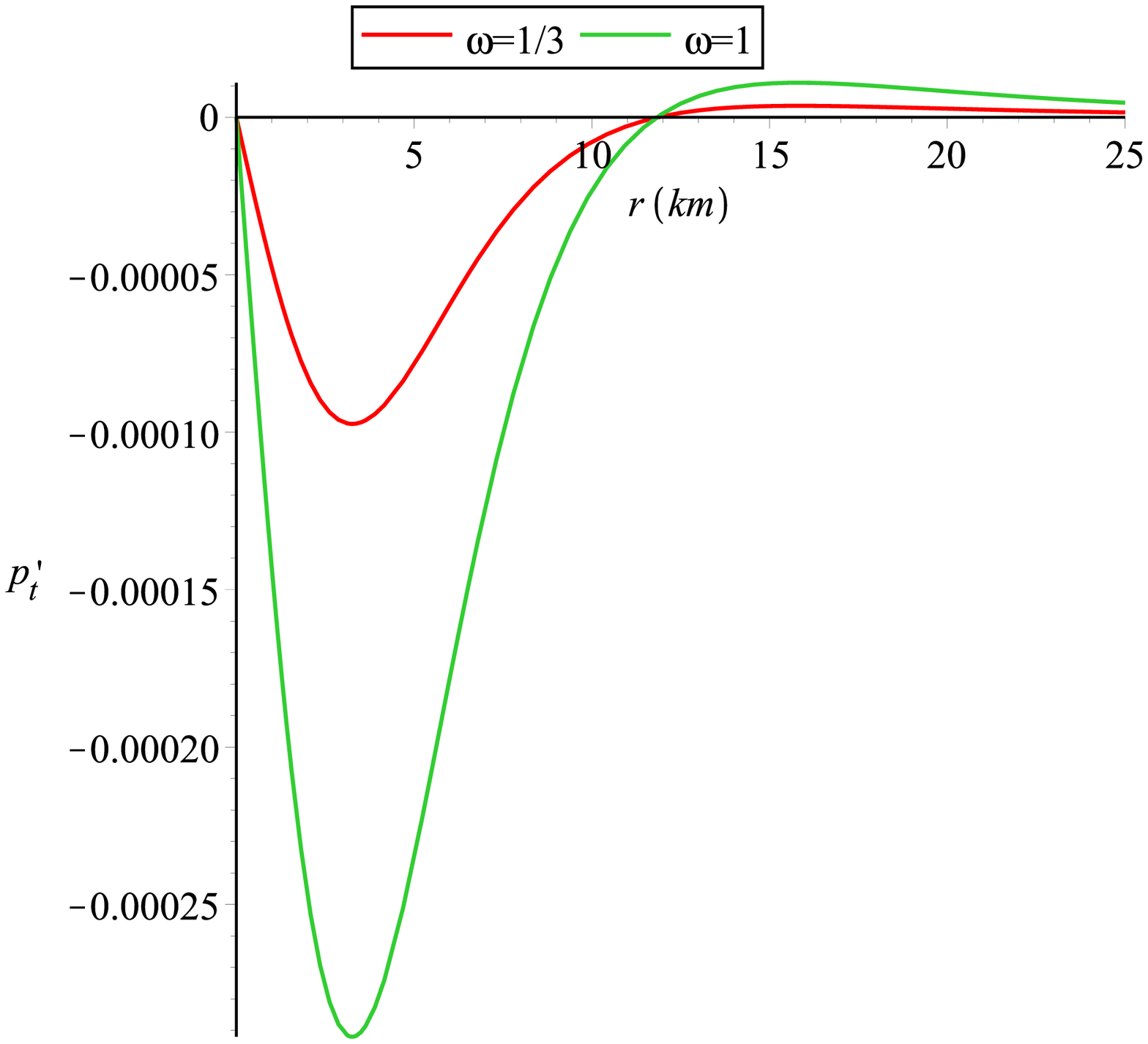}
\includegraphics[width=0.35\textwidth]{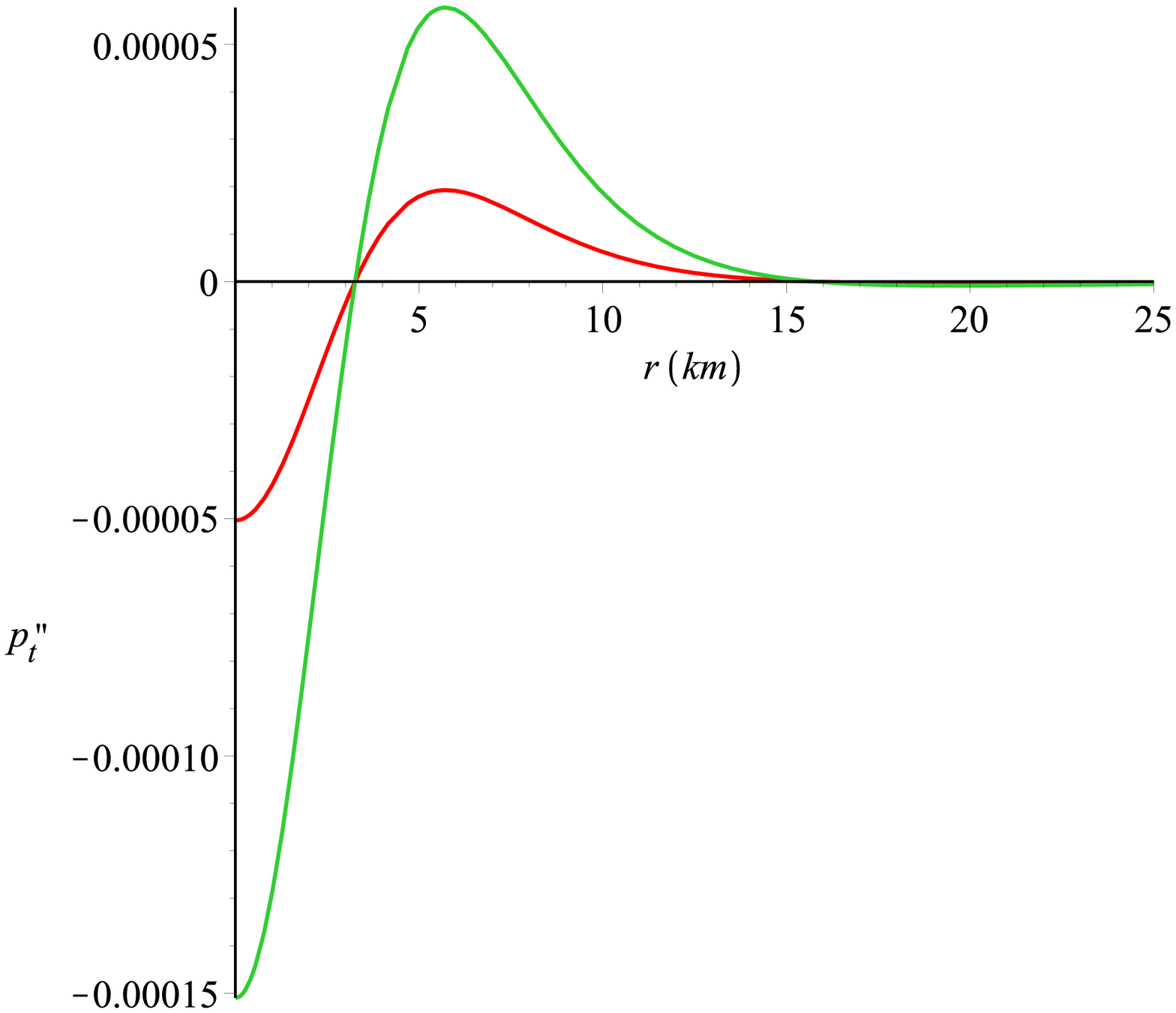}
\caption{Variation of the $p_t$ (upper left), $p_t^{'}$ (upper right) and $p_t^{''}$ (lower) with $r$}
\end{figure*}

Form Fig. 3 it can be seen that the tangential pressure
decreases with the radius and attain a minimum value which then
again increases. This clearly indicates about pulsating nature of
the compact star.

The central density is given by
\begin{equation}
\rho(0)=\frac{1}{2\pi}\left[\frac{(a-1)}{R^2}-\Lambda\right]. \label{eq22}
\end{equation}

The central radial pressure is given by
\begin{equation}
p_r(0)=\omega\left(\frac{1}{2\pi}\left[\frac{(a-1)}{R^2}-\Lambda\right]\right)=\omega\rho(0). \label{eq23}
\end{equation}
The central tangential pressure is given by
\begin{equation}
p_t(0)=\frac{1}{2\pi R^2}(A-B)+\frac{\Lambda}{2\pi}. \label{eq24}
\end{equation}

From the above expressions we note that both the central density
and central pressure are finite at the center of the star. So the present
model is free from any central singularity.

\section {Matching Condition}
The exterior ($p=\rho=0$) solution corresponds to a static,
BTZ-type black hole is written in the following form as
\begin{eqnarray}
ds ^2 = -\left(-M_0 - \Lambda r^2  \right) dt^2 + \left(-M_0 -
\Lambda r^2  \right)^{-1} dr^2 \nonumber \\+ r^2 d\theta^2, \label{eq21}
\end{eqnarray}
where $M_0$ is the conserved mass of the black hole which is
associated with asymptotic invariance under time-displacements.
Here we match the interior spacetime to the BTZ exterior
at the boundary outside the event horizon. Continuity of the metric
functions $g_{tt}$ and $g_{rr}$ at $r=b$, radius of the compact
star, gives the value of the integration constant of Eq.~(\ref{eq14}) as
\begin{equation}
D=\frac{1}{2}(\omega+1)\left[\ln\left(\frac{R^2+b^2}{R^2+a
b^2}\right)+\Lambda (a b^2-R^2\ln a )\right]. \label{eq32}
\end{equation}

Putting $\omega=1/3$ we get the value of integration constant
as $D=-0.5638$ and for $\omega=1$ it yields as $D=-0.8457$.

\section {Physical Analysis}

\subsection{Anisotropic Behavior}
For the model under consideration the measure of anisotropy in
pressure can be obtained as
\begin{equation}
\Delta \equiv (p_t-p_r). \label{eq25}
\end{equation}

It can be seen that the 'anisotropy' will be directed  outward
when $p_t>p_r$ i.e. $\Delta>0$ and inward when $p_t<p_r$ i.e.
$\Delta<0$.

\begin{figure*}[thtb]
\centering
  \includegraphics[width=0.35\textwidth]{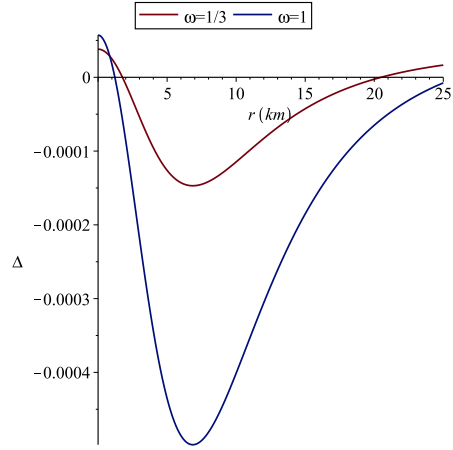}
  \caption{Variation of the anisotropic factor  $\Delta$ with $r$}
\end{figure*}

The profile of $\Delta$ with the radial distance is shown in Fig. 4.
 From this figure it is clear that the anisotropic factor does
not vanish at the center of the star. It is positive from $1$ km
(for $\omega=1$) to $2$ km (for $\omega=1/3$) i.e. $p_t>p_r$. This
implies anisotropy is repulsive. Again for $\omega=1$ the
anisotropic factor is negative between $1$ km to $25$ km and for
$\omega=1/3$ the anisotropic factor is negative between $2$ km to
$20$ km. After that $\Delta$ again increases to positive value.
Since for the maximum part of the stellar distribution the
anisotropy is negative so this allows construction of a more
massive stellar structure as shown by~\citet{Ray2012}.

\subsection{Energy Condition}
For an anisotropic fluid sphere the energy conditions, viz. Weak
Energy Condition (WEC), Null Energy Condition (NEC), Strong Energy
Condition (SEC) and Dominant Energy Condition (DEC) are satisfied
if and only if the following inequalities hold simultaneously by
every points inside the fluid sphere:

$NEC: \rho+p_r \geq 0$,

$WEC: \rho+p_r \geq 0,  \rho> 0$,

$SEC: \rho+p_r \geq 0,  \rho+p_r+2p_t> 0$,

$DEC: \rho>|{p_r}|,  \rho>|{p_t}|$.

We have shown the above inequalities by the help of graphical
representation.

\begin{figure*}[thbp]
\centering
\includegraphics[width=0.35\textwidth]{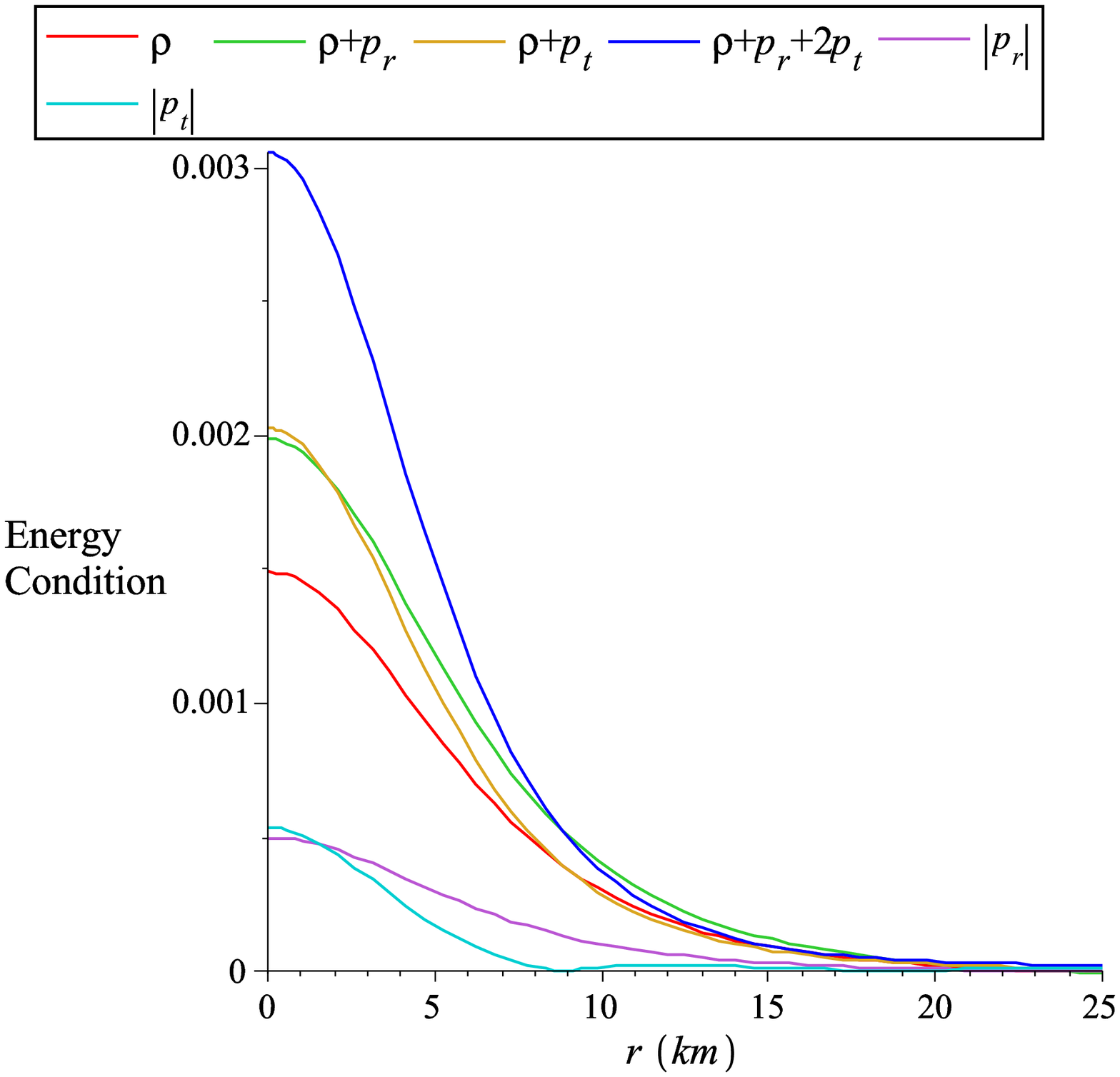}
\includegraphics[width=0.35\textwidth]{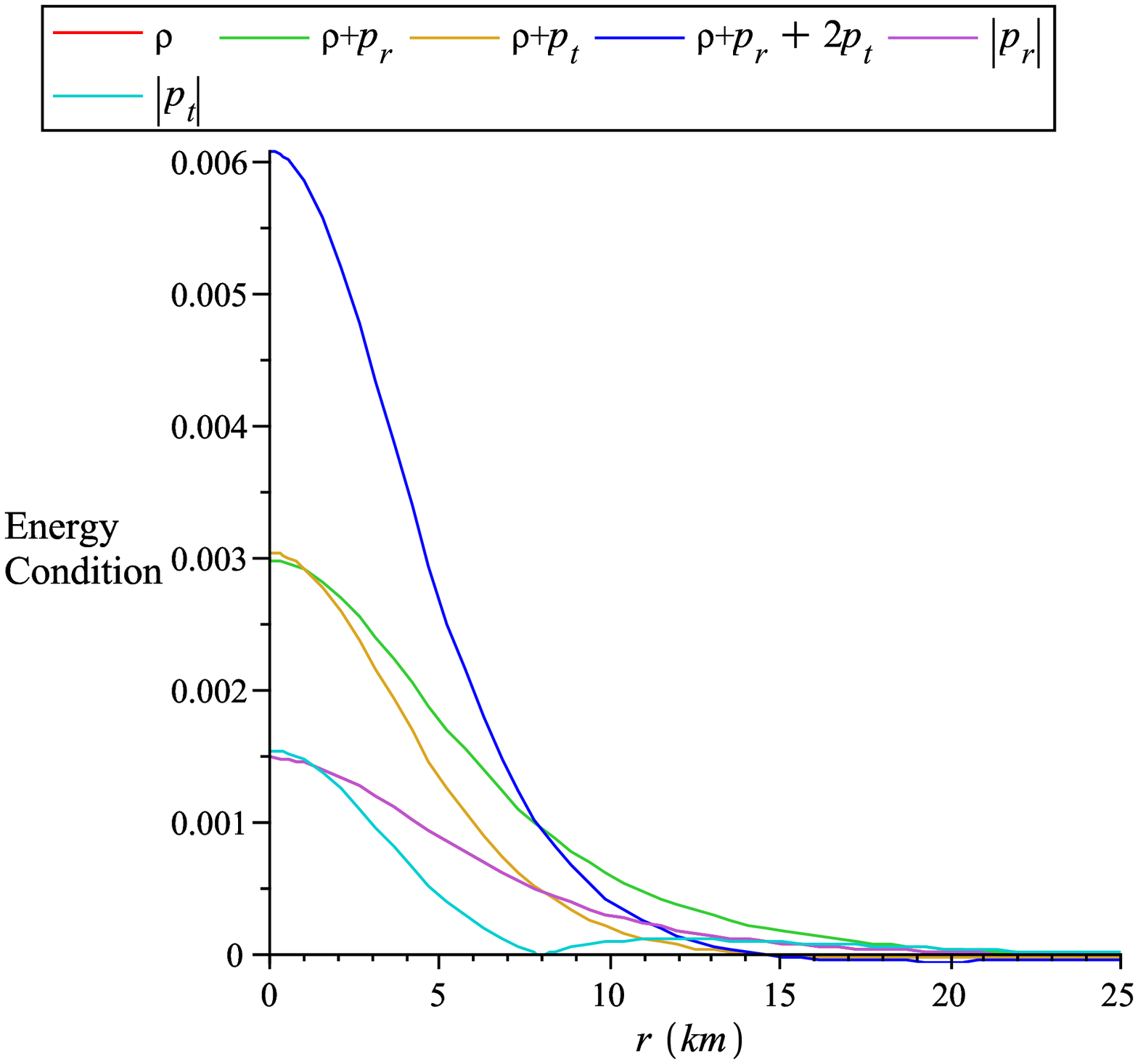}
\caption{Variation of the Energy Condition against $r$ for  $\omega=\frac{1}{3}$ (left panel) and $\omega=1$(right panel)}
\end{figure*}

Fig. 5 shows the energy condition for $\omega=1/3$ (left panel). In this
representation all the energy conditions are satisfied for our
model. On the other hand, right panel of Fig. 5 shows the variation for $\omega=1$. This
variation also shows that our model is satisfied for all the energy
condition and our model provides a stable stellar configuration.
However, as the graphs for $\rho$ and $p_r$ do overlap so
we observe only five graphs.

\subsection{Stability}
The velocity of sound should follow the condition $0<v_s^{2}=
dp/d\rho <1$ for a physically realistic
model~\citep{Herrera1992,Abreu2007,Karar2012}. This condition is
known as causality condition. For our anisotropic model, the
radial and transverse velocities of sound are defined by
\begin{equation}
v_{rs}^{2}=\frac{dp_r}{d\rho}=\omega, \label{eq30}
\end{equation}

\begin{equation}
v_{ts}^{2}=\frac{dp_t}{d\rho}. \label{eq31}
\end{equation}

Due to the mathematical complexity of the expression for
$v_{ts}^{2}$ we shall show the inequality with the help of graphical
representation only.

\begin{figure*}[thbp]
\centering
\includegraphics[width=0.35\textwidth]{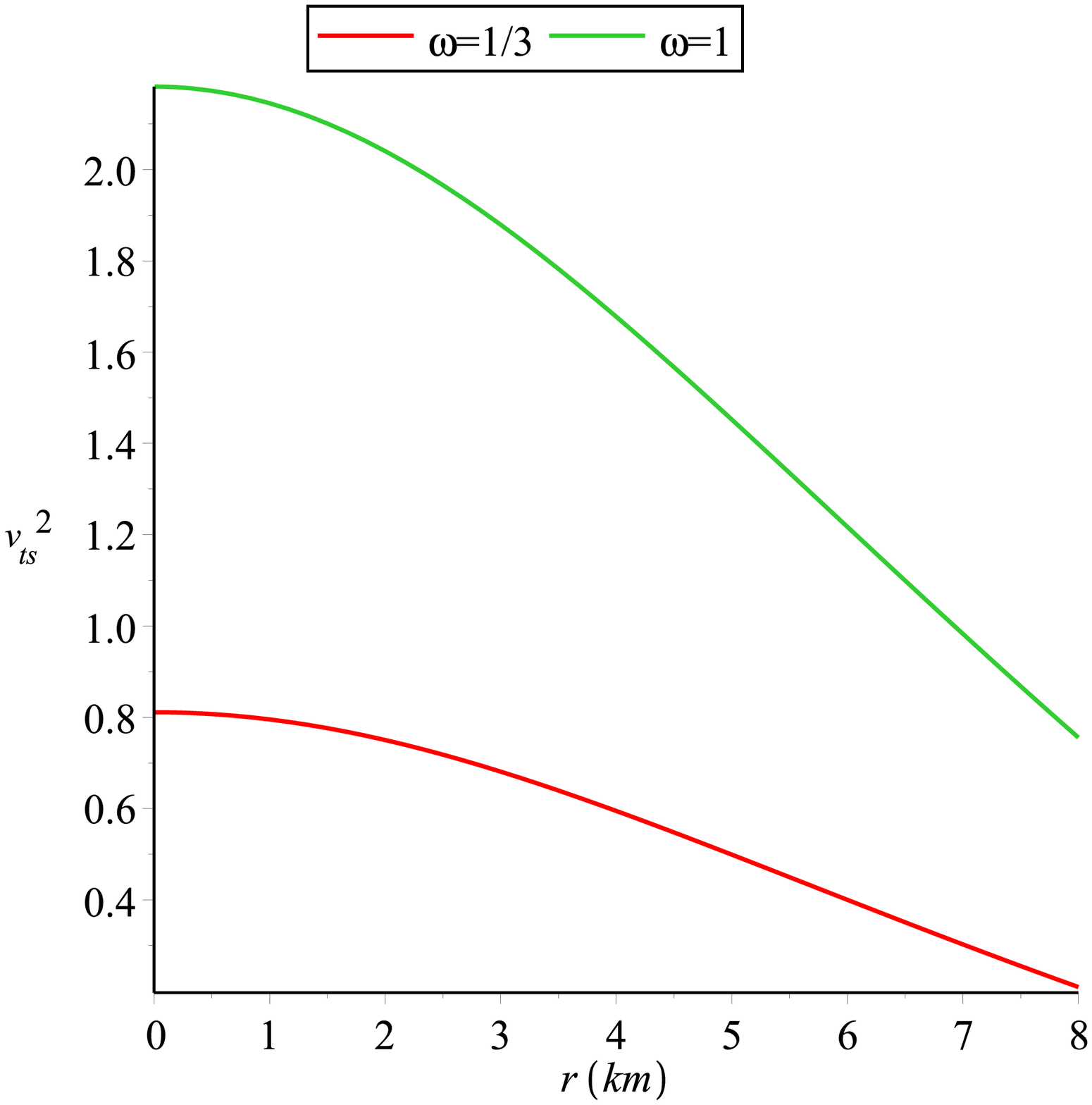}
\includegraphics[width=0.35\textwidth]{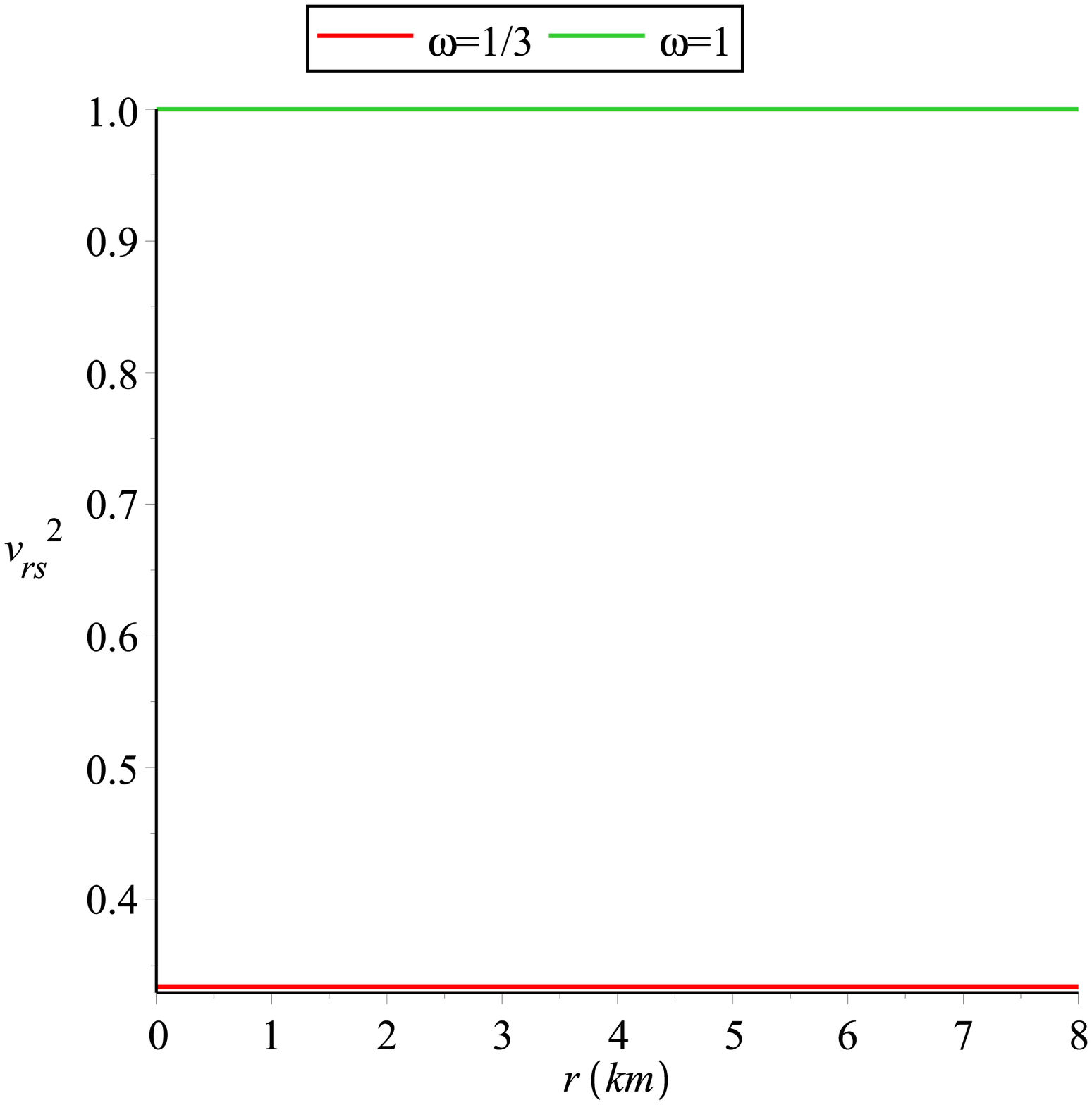}
\includegraphics[width=0.35\textwidth]{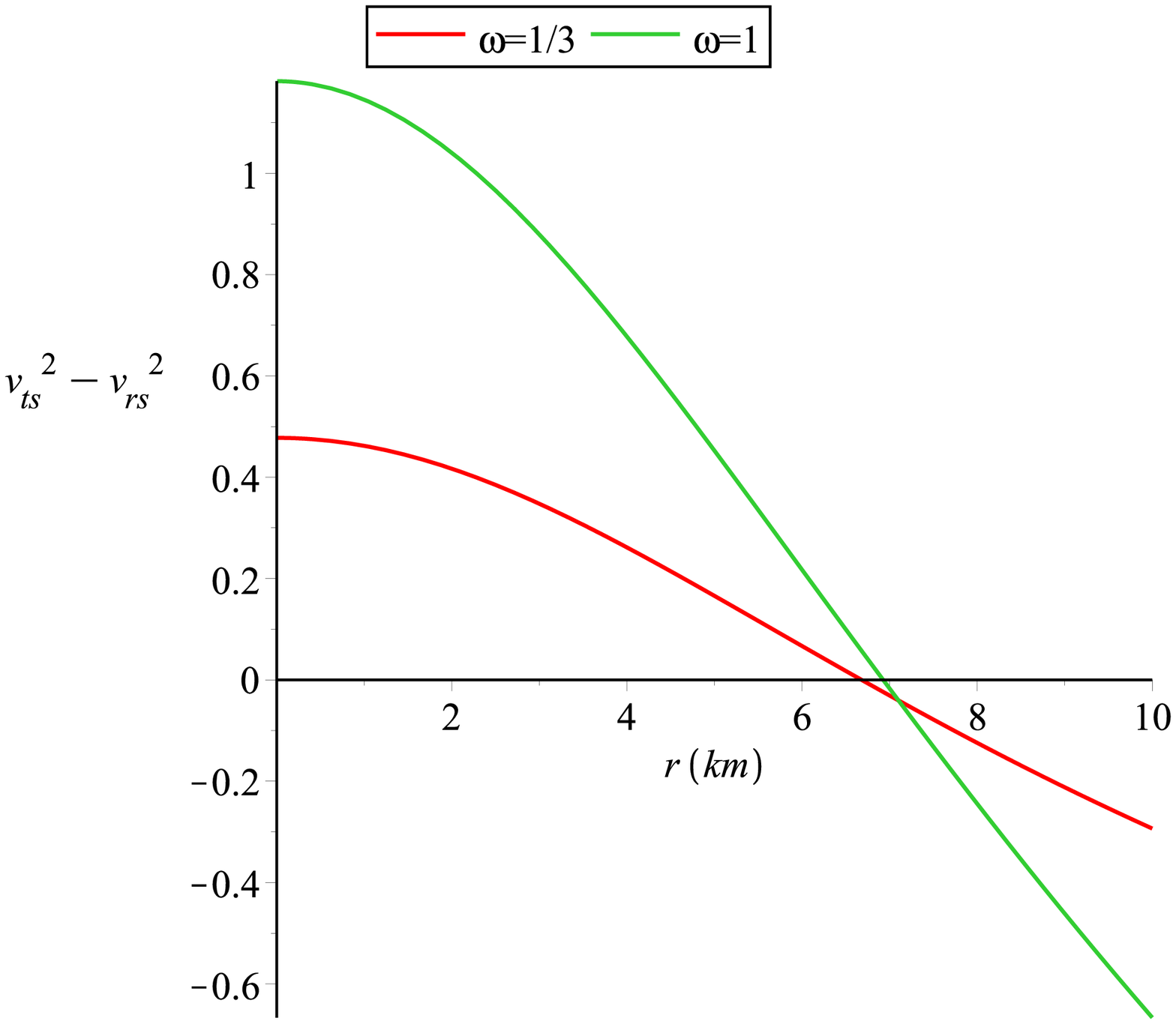}
\caption{Variation of the transverse sound velocity $v_{ts}^2$ (upper left),
radial sound velocity $v_{rs}^2$ (upper right) and $v_{ts}^2-v_{rs}^2$ (lower) with $r$}
\end{figure*}

The variation of $v_{ts}^2$, $v_{rs}^2$ and $v_{ts}^2-v_{rs}^2$
with the radial distance $r$ are shown in Fig. 6.
Herrera proposed a technique for stability check of local anisotropic
  matter distribution. This technique is known as the cracking
concept which states that the region for which radial speed of sound is greater than the
transverse speed of sound is a potentially stable region. Fig. 6
 indicates that there is a change of sign for the term
$v_{ts}^2-v_{rs}^2$ within the specific configuration and thus
confirming that the model has a transition from unstable to stable
configuration. The present stellar model gradually gets stability
with the increase of the radius.

\subsection{Buchdahl condition}
The mass of the compact star can be calculated from the density
profile
\begin{eqnarray}
 m(r) = \int 4\pi r'^2 \rho dr \nonumber\\
= 2 R^2(a-1)\left[-\frac{r}{2a(R^2+ar^2)} \right. \nonumber\\ \left.  +\frac{\arctan(\frac{\sqrt{a} r}{R})}{2a^\frac{3}{2}R}\right]
-\frac{2\Lambda r^3}{3}. \label{eq33}
\end{eqnarray}
Now the mass function is regular at the origin as $r\rightarrow 0$
$m(r)\rightarrow 0$. The profile of mass function is depicted in
Fig. 7. From this figure it is clear that the mass function is
monotonic increasing function of r and for $0\leq r \leq b$
$m(r)>0$.

\begin{figure*}[thtb]
\centering
  \includegraphics[width=0.35\textwidth]{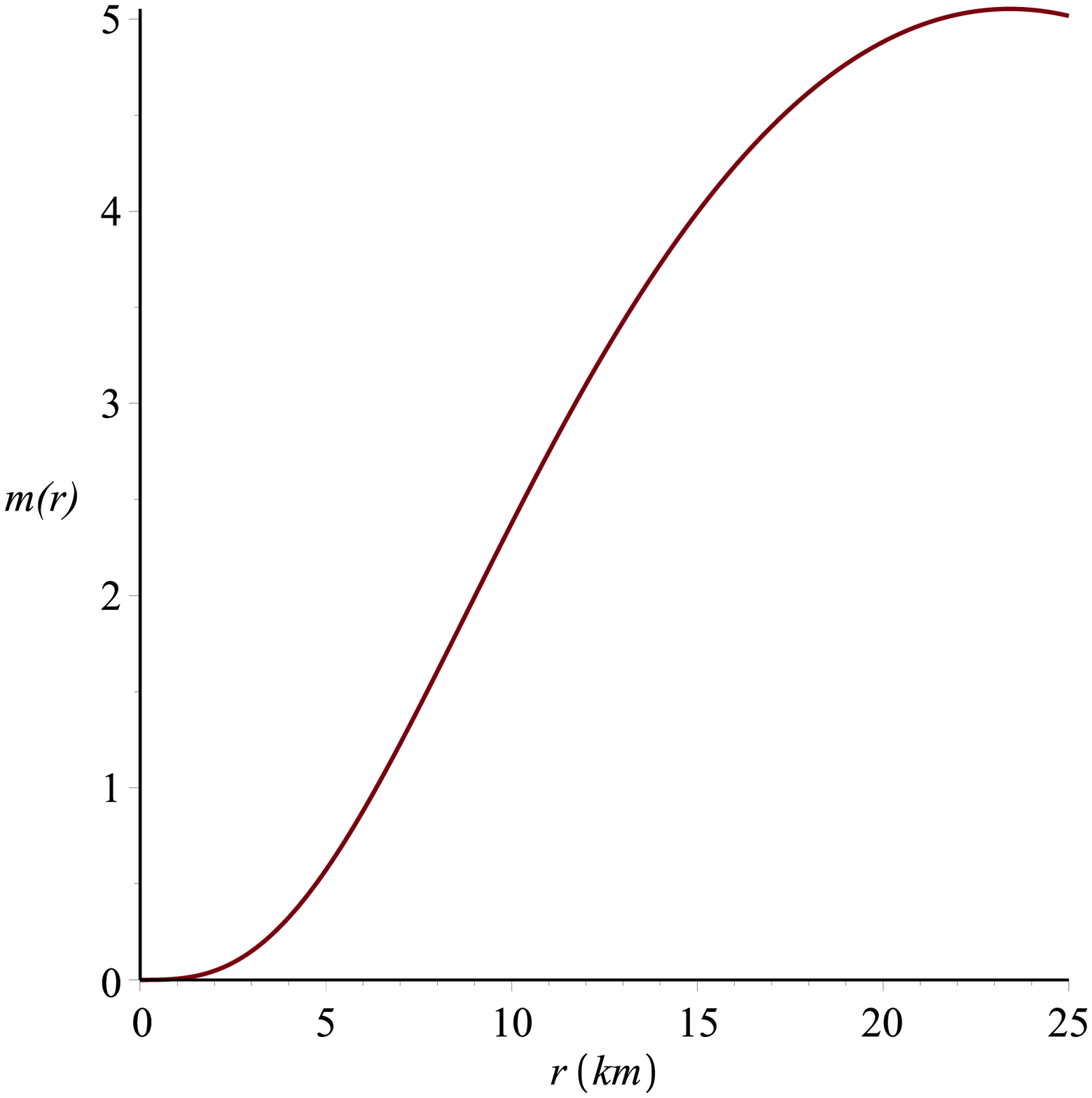}
  \caption{variation of the mass $m(r)$ with the radial distance $r$}
\end{figure*}

The maximum allowable ratio of the mass to the radius of a compact
star can not be arbitrarily large. According to~\citet{Buchdahl1959}
 the ratio of twice the maximum allowable mass to the radius is less than $8/9$, i.e. $2M/R<8/9$ where $M/R$
is called the compactification factor which classifies the stellar
objects in different categories as given by~\citet{Jotania2006} as follows: (i) For normal star
$M/R\sim 10^{-5}$, (ii) For white dwarf $M/R\sim 10^{-3}$, (iii)
For neutron star $10^{-1}<M/R<1/4$, (iv) For ultracompact star
$1/4<M/R<1/2$ and (v) For black hole $M/R=1/2$.

The compactification factor of our model is given by
\begin {eqnarray}
u(r)&=&\frac{m(r)}{r} \nonumber\\
&=&\frac{R(a-1)\arctan(\frac{\sqrt{a}
r}{R})}{a^\frac{3}{2}r}-\frac{R^2(a-1)}{a(R^2+ar^2)} -\frac{2\Lambda
r^2}{3}. \nonumber\\ \qquad \label{eq33}
\end{eqnarray}

\begin{figure}[!htb]
\centering
  \includegraphics[width=0.35\textwidth]{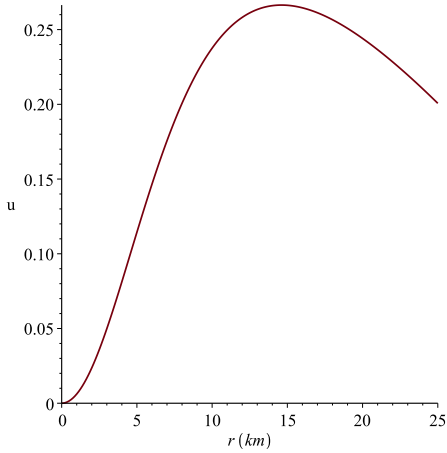}
   \caption{Variation of the compactification  factor  $u$ with the radial distance $r$}
  \end{figure}

The variation of the mass and compactification factor with the
radius of the star are shown in Figs. 7 and 8 respectively both of which are monotonic
increasing function of $r$. Specifically, from the
range of the compactification factor with its maximum value $>0.25$ of Fig. 8 we can
conclude that our model of anisotropic star is an ultracompact
star. We also calculate the redshift of our model
\begin{eqnarray}
Z_s=(1-2u)^{-\frac{1}{2}}-1 \nonumber\\ =\left[1-\frac{2R(a-1)\arctan(\frac{\sqrt{a}
r}{R})}{a^\frac{3}{2}r} \right. \nonumber\\ \left. \qquad+\frac{2R^2(a-1)}{a(R^2+ar^2)} +\frac{4\Lambda
r^2}{3}\right]^{-\frac{1}{2}}-1. \label{eq34}
\end{eqnarray}

\begin{figure}[!htb]
\centering
  \includegraphics[width=0.35\textwidth]{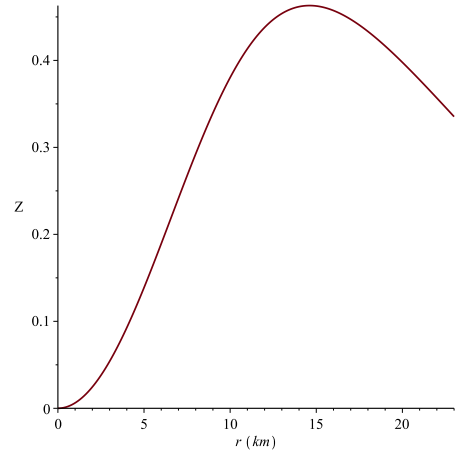}
  \caption{Variation of the redshift $Z$ with the radial distance $r$}
  \end{figure}

The profile of the redshift function of our compact star is shown
in Fig. 9. In this connection we want to mention that for
anisotropic star the value of the maximum surface redshift for
our model is near about $0.3$.

\section{Concluding remarks}
We have studied in the present work a $(2+1)$ dimensional compact
star in the pseudo spheroidal spacetime. The motivation behind the
study is the consideration that (i) the three-dimensional black
holes are interesting by themselves, and (ii) the toy models in
$3D$ may have a radically different qualitative behavior with
respect to the more realistic $4D$ setups. Under this background
we have studied the interior solution of a $(2+1)$ dimensional
compact star in the pseudo spheroidal spacetime. The main
guideline used in this work is a metric found in the Ref.
\citep{Vaidya1982}, where a spheroidal 3-space exhibits central
symmetry. It is clear that the investigation of the field of
spheroidal bodies in general relativity has important
astrophysical consequences, due to the fact the rotation of
planets and stars produce spheroidal shapes. So, it seems that is
a good approximation (slow rotation) to consider an static
spheroidal spacetime to represent the gravitational field of these
celescial bodies.

Therefore in the present work the Vaidya-Tikekar prescription
\citep{Vaidya1982} has been employed to get the matter density,
radial pressure and other quantities from the Einstein field
equations. Some salient and interesting features of the study can
be put forward as follows:

(1) The matter density and radial pressure both are regular at the center
and are monotonic decreasing function of the radial parameter. This
behavior indicate that they have maximum value at the center and it
decreases from the center to the boundary of the star. At the
boundary of the star the radial pressure and matter density do vanish
as expected.

(2) The tangential pressure is also a decreasing function of the radial
distance. It decreases rapidly and within the stellar structure it
has a fluctuating nature.

(3) The metric functions $g_{rr}$ and $g_{tt}$ are continuous at
the boundary of the star. From this situation one can calculate the
value of integration constant $D$.

(4) It is well known that the anisotropic factor $\triangle$
should vanish at the origin but our model does not show this feature
rather it has a finite positive value. However, for the maximum part of our
stellar distribution the anisotropy is repulsive which allows
formation of more massive star.

(5) Our model satisfies all the energy conditions for $\omega=1/3$
and $\omega=1$. So our model provides a stable stellar configuration.

(6) We observe that by obeying Herrera's cracking condition our
model maintains stabilty with increase of the radius.

(7) The mass, compactification factor and surface redshift all are
monotonic increasing function of the radius of the star. The maximum
value of the compactification factor indicates that our model represents
an ultra compact object. The maximum value of the surface redshift is about $0.3$
for the present model.

Finally, we would like to made some comments on our toy models in 3-dimension
relative to the 4-dimensional one. As mentioned earlier, it seems that the 3D may have
a radically different qualitative behavior with respect to the 4D having a more realistic setups.
In connection to this it is to be noted that we have used the data from a
3-spatial dimensional object in order to estimate the constants of the model.
 Even though one can not have a strong argument about the physical meaning
of the consideration of stellar objects of 2-spatial dimensions, however,
 for an observer in the plane, $\theta = constant$, all characteristics will reveal as a
$(2+1)$ dimensional portrait. So superficially it will be more or
less justified to use all the data which are apparently the same
for both the spacetime. However, regarding development of a valid
approach to represent lower dimensional gravity with spheroidal
shapes one may follow the methodology of the Refs. \citep{Quevedo1989,Chifu2012}
for further investigation.

\section*{Acknowledgement}
FR and SR thank the Inter-University Center for Astronomy and Astrophysics
(IUCAA), Pune for providing the Associateship programme which has facilitated
to start working on the problem. Also SR is thankful to the authority of
The Institute of Mathematical Sciences, Chennai, India
for providing Associateship under which a part of this work was carried out.

\end{document}